\documentclass[a4]{article}

\usepackage{epsfig}
\usepackage{amsmath}
\allowdisplaybreaks

\usepackage{thophys}
\usepackage{thohacks}

\begin{document}
\setlength{\unitlength}{1mm}
\renewcommand{\thefootnote}{\fnsymbol{footnote}}
\thispagestyle{empty}
\begin{titlepage}
\begin{flushright}
hep-ph/9801436 \\
TTP 98--02 \\
\today
\end{flushright}

\vspace{0.3cm}
\boldmath
\begin{center}
\Large\bf  Renormalizing Heavy Quark Effective Theory at $\mathcal O(1/m_{Q}^{3})$
\end{center}
\unboldmath
\vspace{0.8cm}

\begin{center}
{\large Christopher Balzereit\footnote[5]{email:
    chb@particle.physik.uni-karlsruhe.de}}\\
{\sl Institut f\"{u}r Theoretische Teilchenphysik,
     Universit\"at Karlsruhe,\\ D -- 76128 Karlsruhe, Germany} 
\end{center}

\vspace{\fill}

\begin{abstract}
\noindent
We present a calculation of the renormalized HQET Lagrangian
at order $\mathcal O(1/m_Q^3)$ in the one particle sector. 
The anomalous dimensions 
of local operators and
time ordered products of dimension 7 contributing at this
order are calculated in the one loop approximation.
We show that a careful treatment of the time ordered
products is necessary to arrive at a gauge independent 
renormalized lagrangian. 
Our result sets the stage for an investigation of reparametrization
invariance at $\mathcal O(1/m_Q^3)$.

\end{abstract}
\end{titlepage}

\renewcommand{\thefootnote}{\arabic{footnote}}

\newpage

\section{Introduction}
\label{sec:introduction}
Heavy Quark Effective Field Theory
(HQET)~\cite{HQET_classic} has been
established as the theoretical tool of choice for the description of
mesons and baryons containing one heavy
quark~\cite{HQET_review}.  This derives from the fact that it is a
systematic expansion in inverse powers of the heavy quark mass with
well defined and calculable coefficients.  Furthermore, its
realization of the spin and flavor symmetry of the low energy theory
is a phenomenologically powerful tool.

The $1/m_Q$ expansion has already been applied sucessfully to 
phenomenological problems such as the determination of $V_{cb}$. 
This involves a matrix element of the left handed current for the 
$b \to c $ transition between two heavy hadrons. In order to 
match the experimental precision corrections ($1/m_Q$ as well 
as radiative corrections) are indispensable. 

The present paper considers partial aspects of the $1/m_Q^3$  
corrections, which eventually are needed for phenomenological 
applications. Here the lagrangian of QCD is expanded up to order
$1/m_Q^3$ and the one-loop renormalization of the corresponding 
operators is calculated. Results for the lower order terms 
\cite{eichten,grozin,balzi,blok,bauer,finke} 
as well as a matching calculation for terms of order $1/m_Q^3$ 
\cite{manohar} are already known, but the full renormalization has not 
yet been studied. 

The lagrangian itself is also of theoretical interest, since on 
one side it is relevant for the $1/m_Q$ expansion of the heavy 
hadron mass, and on the other side allows a study of 
reparametrization invariance. 



 

To keep things as simple as possible we restrict
ourselves to the one particle sector of
HQET, disregarding four fermion operators as well as 
pure gluonic operators.
The calculation of the anomalous dimensions of 
the corresponding operator basis sets the stage
for an extensive study of reparametrization
invariance at $\mathcal O(1/m_{Q}^{3})$, which
will be subject of a sequel to this paper.

This note is organized as follows: in section~\ref{sec:basis} we introduce
our operator basis and discuss its reduction
to a set of linearly independent operators.  
Our result for the anomalous dimensions 
and some checks of its consistency are presented
in section~\ref{sec:anodim}.
In section~\ref{sec:calculation} we give insight into
some aspects of the calculation.  
Finally, we
present the logarithmic contributions to the short distance coefficients 
of the effective lagrangian
in section~\ref{sec:RG-flow} and
conclusions in section~\ref{sec:conclusions}.

\section{Operator basis}
\label{sec:basis}

The effective HQET lagrangian is defined by a systematic expansion
of QCD in inverse powers of the heavy quark mass
\begin{equation} 
\label{eq:HQETlag}
  \mathcal{L}_{\text{HQET}} = \bar h_{v}(i v D) h_{v}
    +\sum_{n=1}^{\infty}\frac{1}{(2m_Q)^n} \sum_i  C_i^{(n)} 
     \mathcal{O}_i^{(n)} \,.
\end{equation}
To lowest order only one operator, $\bar h_{v}(i v D) h_{v}$, contributes,
which is independent of the flavor and the spin of the heavy quark,
resulting in the well known spin-flavor symmetry of HQET.
However, these symmetries are valid only in the
limit $m_Q \rightarrow \infty$ and broken by local operators 
$\mathcal{O}_i^{(n)}$ of dimension 5 and higher.
The coefficients $C_i^{(n)}$ are short distance
corrections, which compensate for the modified UV behavior
of the effective theory. They
are determined perturbatively by matching
HQET to QCD order by order in the
strong coupling $\alpha_s$ and the inverse heavy quark mass
supplemented by a renormalization group analysis of the respective operators.\\
 
We start with a review of the local operators
appearing at $\mathcal O(1/m_Q)$ and $\mathcal O(1/m_Q^2)$,
respectively. 
At $\mathcal O(1/m_Q)$ we choose the conventional basis
\begin{equation}
\begin{aligned}
  \mathcal{O}^{(1)}_1
    &= \bar h_v (iD)^2 h_v \\
  \mathcal{O}^{(1)}_2
    &= \frac{g}{2} \bar h_v \sigma^{\mu \lambda} 
       F_{\mu \lambda} h_v \\
  \mathcal{O}^{(1)}_3
    &= \bar h_v (ivD)^2 h_v 
\end{aligned}
\end{equation}
and at $\mathcal O(1/m_Q^2)$
\begin{equation}
\begin{aligned}
  \mathcal{O}^{(2)}_1
    &= \bar h_v iD_{\mu} (ivD) iD^{\mu} h_v &\qquad
  \mathcal{O}^{(2)}_2
    &= \bar h_v i \sigma^{\mu \lambda} iD_{\mu}
       (ivD) iD_{\lambda} h_v  \\
  \mathcal{O}^{(2)}_3
    &= \bar h_v (ivD) (iD)^2 h_v &\qquad
  \mathcal{O}^{(2)}_4
    &= \bar h_v (iD)^2 (ivD) h_v  \\
  \mathcal{O}^{(2)}_5
    &= \bar h_v (ivD)^3 h_v &\qquad
  \mathcal{O}^{(2)}_6
    &= \bar h_v (ivD) i \sigma^{\mu \lambda} iD_{\mu}iD_{\lambda}
       h_v  \\
  \mathcal{O}^{(2)}_7
    &= \bar h_v i \sigma^{\mu \lambda}
       iD_{\mu} iD_{\lambda} (ivD) h_v \,. & 
\end{aligned}
\end{equation}
The definition of the covariant derivative 
$iD = i\partial + g_sT^{a}A^{a}$ and field strength tensor
$F_{\mu\nu}^{a}T^{a} = -i/g_{s} [iD_{\mu},iD_{\nu}]$ follows usual conventions.
Note, that we refrain from skipping  operators which
vanish by the heavy quark equation of motion (EOM)
$(ivD)h_v=0$. This will
proove important, since once inserted into  
time ordered products at $\mathcal O(1/m_Q^3)$, such
operators contribute to physical operators,
as will be explained in detail below.

At  $\mathcal O(1/m_Q^3)$ we find 13 local operators
contributing to physical matrixelements:
\begin{equation}
\begin{aligned}
\mathcal{O}^{(3)}_{1} &= \bar h_v iD_{\mu}(ivD)^2 iD^{\mu} h_v &\qquad
\mathcal O^{(3)}_{2} &= \bar h_v (iD)^2(iD)^2 h_v \\
\mathcal O^{(3)}_{3} &= \bar h_v iD_{\mu}(iD)^2 iD^{\mu}h_v &\qquad
\mathcal O^{(3)}_{4} &= \bar h_v iD_{\mu}iD_{\nu} iD^{\mu}iD^{\nu}h_v \\
\mathcal O^{(3)}_{5} &= \bar h_v i\sigma^{\mu\nu}iD_{\mu} (ivD)^2 iD_{\nu} h_v  &\qquad
\mathcal O^{(3)}_{6} &= \bar h_v i\sigma^{\mu\nu}iD_{\mu}iD_{\nu}(iD)^2 h_v \\
\mathcal O^{(3)}_{7} &= \bar h_v i\sigma^{\mu\nu}iD_{\rho}iD_{\mu}iD_{\nu}iD^{\rho}h_v &\qquad
\mathcal O^{(3)}_{8} &= \bar h_v i\sigma^{\mu\nu}(iD)^2 iD_{\mu}iD_{\nu}h_v \\
\mathcal O^{(3)}_{9} &= \bar h_v i\sigma^{\mu\nu}iD_{\mu}(iD)^2iD_{\nu}h_v &\qquad
\mathcal O^{(3)}_{10} &= \bar h_v i\sigma^{\mu\nu}iD_{\mu}iD_{\rho}iD_{\nu}iD^{\rho}h_v \\
\mathcal O^{(3)}_{11} &= \bar h_v i\sigma^{\mu\nu}iD_{\rho}iD_{\mu}iD^{\rho}iD_{\nu}h_v &\qquad
\mathcal O^{(3)}_{12} &=g_s^2 \bar h_v F^{a\mu\nu}F^a_{\mu\nu}  h_v \\
\mathcal O^{(3)}_{13} &=g_s^2 \bar h_v v_{\nu}v^{\rho}F^{a\mu\nu}F^a_{\mu\rho}h_v & 
\end{aligned}
\end{equation}
Below these operators will be denoted collectively
by $\vec{\mathcal O}^{(3)}$.

Another 11 operators contain the EOM piece $(ivD)h_v$ and are not 
considered any further, since at fixed order $1/m_Q^{3}$
their matrix elements vanish.   

In addition to the local operators, there are the time-ordered
products of the lower dimensional operators $\mathcal{O}_i^{(1)}$
and $\mathcal{O}_i^{(2)}$.
Their contributions at $\mathcal O(1/m_Q^2)$ and $\mathcal O(1/m_Q^3)$  
are written generically as
\begin{equation}
\mathcal T^{(11)}_{ij} =(1 - \frac{1}{2}\delta_{ij})
 i\Tprod{\mathcal O^{(1)}_i,\mathcal O^{(1)}_j},
\quad i,j = 1\ldots 3, \quad i\leq j
\end{equation}
and 
\begin{align}
\mathcal T^{(12)}_{ij}&=
i\Tprod{\mathcal O^{(1)}_{i},\mathcal O^{(2)}_{j}},\quad i = 1\ldots 3, \quad j =
1\ldots 7\\
\mathcal T^{(111)}_{ijk}&=-\mathcal S_{ijk}
\Tprod{\mathcal O^{(1)}_{i},\mathcal O^{(1)}_{j},\mathcal O^{(1)}_{k}},
\quad i,j,k = 1\ldots 3,\quad i\leq j \leq k \,.
\end{align}
The symmetry factor $\mathcal S_{ijk}$ equals $1 ,1/2$ or
$1/6$, if no, two or all inserted operators are identical.
There is a  total of 31 time ordered products
but not all of them are physical.
The reason is that time ordered products, which contain at 
least one operator vanishing by the
EOM, can contract to local operators if the $(ivD)h_v$ term
acts on an internal heavy quark line. 
This is  illustrated graphically as
follows: 

\leavevmode
\epsfxsize=9cm
\epsffile[180 640 470 700]{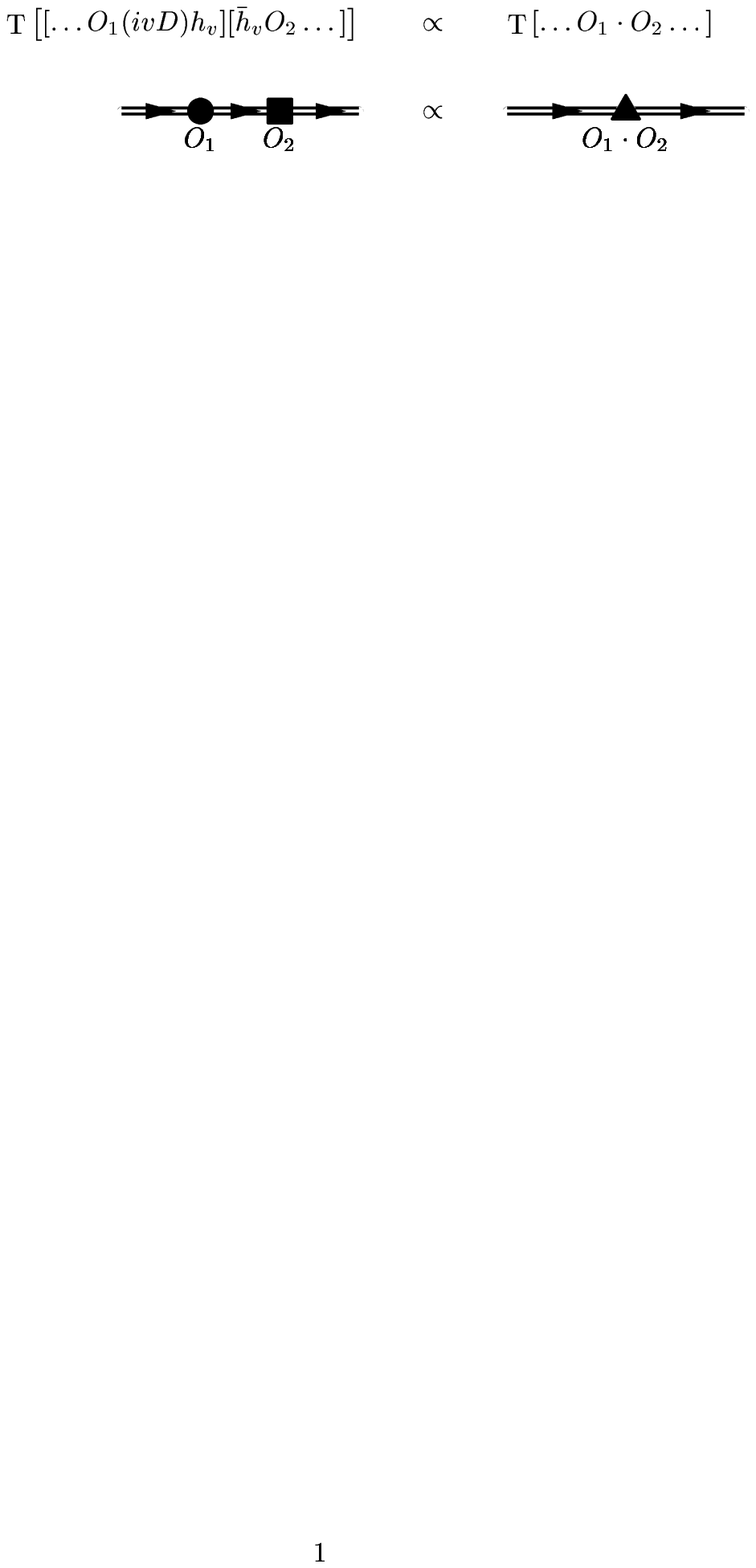} 

Below such operator relations will be denoted 
{\it contraction identities} (CI). They can be derived by manipulating
the generating functional of Greensfunctions in the presence
of the relevant operators
in much the same way, as one usually establishes the
validity of the EOM for local operators.
As a concrete example consider 
the triple insertion 
\begin{equation}
\mathcal T^{(111)}_{113}=-\frac{1}{2}\Tprod{ \bar h_v (iD)^2 h_v,
                        \bar h_v (ivD)^2 h_v,\bar h_v (iD)^2 h_v}.
\end{equation}
Naively acting with the EOM operator on the heavy quark propagators 
to the left and the right, one would
expect that
\begin{equation}
\mathcal T^{(111)}_{113}
= \bar h_v (iD)^2(iD)^2 h_v + \ldots.
\end{equation}
On the other hand an exact derivation of the contraction identity leads to
\begin{equation}
\label{eq:CI111}
\begin{align}
\mathcal T^{(111)}_{113}
&= -i\Tprod{\bar h_v (ivD)(iD)^2 h_v,\bar h_v (iD)^2 h_v} 
-i\Tprod{\bar h_v (iD)^2(ivD) h_v,\bar h_v (iD)^2 h_v} \\
&\quad -  \bar h_v (iD)^2(iD)^2 h_v.\nonumber
\end{align}
\end{equation}
The apparent contradiction in the sign of the local operator 
is resolved by the fact
that the double insertions on the right hand side themselves
contain a local contribution proportional to $\bar h_v (iD)^2(iD)^2 h_v$:
\begin{equation}
\label{eq:CI12}
 i\Tprod{\bar h_v (ivD)(iD)^2 h_v,\bar h_v (iD)^2 h_v} = 
-\bar h_v (iD)^2(iD)^2 h_v +[\mathrm{EOM}]
\end{equation}
Here the symbol $[\mathrm{EOM}]$ collects time ordered
products that vanish by the EOM and can be neclected.
The CI allow us to remove all time ordered
products from the operator basis that contain
EOM operator components. That way we are left with 4 triple insertions
and 4 double insertions collectively denoted as
 $\vec{\mathcal T}^{(111)}$ and $\vec{\mathcal T}^{(12)}$ and composed only of
the lower order physical operators $\mathcal O^{(1)}_{1,2}$
and $\mathcal O^{(2)}_{1,2}$. 
In the next section
we will show explicitly that a consistent implementation of CI 
into the renormalization procedure is necessary to
maintain gauge independence of the renormalized 
lagrangian.\\

\section{Anomalous dimensions}
\label{sec:anodim}

To treat CI correctly during renormalization
we start with the calculation of the anomalous
dimensions corresponding to the {\it full
operator basis}. This basis contains all
local operators of dimension 7 allowed by Lorentz invariance and the
symmetries of HQET as well as all time ordered products
which can be constructed from lower order operators 
disregarding any operator identities (such as CI).
The anomalous dimension matrix corresponding to
the full basis is then projected onto the 
{\it physical basis} with all CI applied and EOM operators removed.
The physical basis is minimal in the sense that it
contains only linearly independent operators.

To clarify this procedure let us for a moment forget about 
time ordered products and consider renormalization
of an overcomplete operator basis consisting 
of physical operators $\mathcal O_i$ and 
operators $\hat{\mathcal O}_j$ which linearely depend on the 
 $\mathcal O_i$:
\begin{equation}
\label{eq:redundant}
\hat{\mathcal O}_i= \sum_{k}c_{ik} \mathcal O_k
\end{equation}
The full basis renormalizes as
\begin{equation}
\begin{align}
\mathcal O_i &= \sum_j Z^{(1)}_{ij} \mathcal O_j^{bare}
             + \sum_j \hat{Z}^{(1)}_{ij} \hat{\mathcal O}_j^{bare}\\
\hat{\mathcal O}_i &= \sum_j Z^{(2)}_{ij} \mathcal O_j^{bare} 
             + \sum_j \hat{Z}^{(2)}_{ij} \hat{\mathcal O}_j^{bare}\,,
\end{align}
\end{equation}
wherein we use (\ref{eq:redundant}) to eleminate the bare 
operators $\hat{\mathcal O}_j^{bare}$ in favor of the physical ones:
\begin{equation}
\label{eq:effren1}
\begin{align}
\mathcal O_i &= \sum_j Z^{(1)}_{ij} \mathcal O_j^{bare} 
             + \sum_{jk} \hat{Z}^{(1)}_{ij} c_{jk} \mathcal O_k^{bare}
              =\sum_j \tilde{Z}^{(1)}_{ij} \mathcal O_j^{bare}\\
\label{eq:effren2}
\hat{\mathcal O}_i &= \sum_j Z^{(2)}_{ij} \mathcal O_j^{bare} 
             + \sum_{jk} \hat{Z}^{(2)}_{ij} c_{jk} \mathcal O_k^{bare}
              =\sum_j \tilde{Z}^{(2)}_{ij} \mathcal O_j^{bare}
\end{align}
\end{equation}
$\tilde{Z}^{(1)}_{ij}$ are effective renormalization 
constants which describe the renormalization of 
the physical operator basis $\mathcal O_i$ among themselves and defines 
their anomalous dimensions.
The $\tilde{Z}^{(2)}_{ij}$ express the renormalization
counterterms of the redundant operators in terms of the
physical ones. They are not linearly independent since 
the renormalized operators on the left hand side
of (\ref{eq:effren1},\ref{eq:effren2}) should also fulfill 
(\ref{eq:redundant}),
which in turn leads to the 
consistency condition
\begin{equation}
\label{eq:consis}
\sum_{k}c_{ik} \tilde{Z}^{(1)}_{kj} = \tilde{Z}^{(2)}_{ij}.
\end{equation}
To one loop order and in the MS scheme, 
the anomalous dimensions are the negative of the
pole parts of the renormalization constants:
\begin{equation}
\tilde{Z}^{(1)}_{ij} = \delta_{ij}  - (\frac{\alpha}{\pi})\frac{1}{\epsilon}
                        \gamma_{ij} + \mathcal O((\frac{\alpha}{\pi})^2)
\end{equation}
Subsequently (\ref{eq:consis}) leads to relations among the 
anomalous dimensions of the physical and unphysical operators,
which provides a powerful consistency check of the
calculation. 

The discussion above applies to our operator basis,
if we identify the $\hat{\mathcal O}_i$ with time ordered products
which are redundant after application of the CI and the $\mathcal O_i$
with the residing physical operators.
In our case the removal of the redundant operators 
prooves important to ensure gauge independence of the physical operator basis.
In the following we demonstrate this explicitly in the case of the renormalized
triple insertion $\mathcal T^{(111)}_{111}$ of the kinetic energy operator $\mathcal O^{(1)}_1$.
Before applying the CI, the renormalized triple insertion expressed  
in terms of bare operators reads 
\begin{equation}
\label{eq:CIexample}
\begin{aligned}
\mathcal T^{(111)}_{111} =\! -(\frac{\alpha}{\pi})\!
                  \frac{1}{\epsilon}\!&\biggl[\quad
 C_F(2+\bar \xi)(-\frac{1}{2})\Tprod{ \bar h_v (iD)^2 h_v,
                        \bar h_v (ivD)^2 h_v,\bar h_v (iD)^2 h_v}^{bare}\\
       &+(\frac{1}{12}C_A+C_F(\frac{10}{3}+\bar\xi))\,i\Tprod{ \bar h_v (iD)^2 (ivD)h_v,\bar h_v (iD)^2 h_v}^{bare}\\
      &   +(\frac{1}{12}C_A+C_F(\frac{10}{3}+\bar\xi))\,i\Tprod{ \bar h_v(ivD) (iD)^2 h_v,\bar h_v (iD)^2 h_v}^{bare}\\
 & +(\frac{1}{6}C_A+C_F(\frac{14}{3}+\bar\xi))\, \bar h_{v}(iD)^{2}(iD)^{2}h_{v}^{bare}
\qquad\qquad\biggr] 
+ \ldots\,.
\end{aligned}
\end{equation}
$\bar \xi = 1 - \xi$ is the usual gauge parameter and the 
ellipses denote terms not relevant
for our discussion.
The first term on the right hand side results from
local renormalization of the kinetic energy operator
\begin{equation}
\bar h_v (iD)^2 h_v =\bar h_v (iD)^2 h_v^{bare} -(\frac{\alpha}{\pi})
  \frac{1}{\epsilon}
 C_F(2+\bar \xi)\bar h_v (ivD)^2 h_v^{bare}\,.
\end{equation}
The fact that the kinetic energy operator is not renormalized 
multiplicatively is a consequence of the celebrated 
reparametrization invariance of HQET \cite{repar}. However, the kinetic
energy operator requires renormalization
by the EOM operator $\mathcal O^{(1)}_{3}$, which in turn finds it
way into renormalization of the triple insertion $\mathcal T^{(111)}_{111}$.

The appearance of the  operators
$\bar h_v (iD)^2 (ivD)h_v$ and $\bar h_v (ivD) (iD)^2 h_v$
in the double time ordered products  on the right hand side of 
(\ref{eq:CIexample}) derives from the fact that
the double insertion of the kinetic energy operator 
needs renormalization by a local operator of 
dimension 6:
\begin{equation}
\label{eq:nlren}
\begin{aligned}
\frac{i}{2}\Tprod{\bar h_v (iD)^2 h_v,\bar h_v (iD)^2 h_v} &= 
-(\frac{\alpha}{\pi})\frac{1}{\epsilon}
\biggl(\frac{1}{12}C_A + C_F(\frac{10}{3} + \bar \xi)\biggr)\\
&\cdot\biggl[\bar h_v(ivD) (iD)^2 h_v^{bare}  +
\bar h_v (iD)^2 (ivD) h_v^{bare} \biggr] + \ldots \, .
\end{aligned}
\end{equation}

If we now replace  the bare triple insertion 
$\mathcal T^{(111)}_{113}$ on the right hand side of 
(\ref{eq:CIexample}) with
the help of CI (\ref{eq:CI111}), 
the gauge dependencies cancel and we are left with 
a gauge independent result. 
We stress that the latter property is not spoiled by
additional application of CI (\ref{eq:CI12}) for the double insertions 
since their coefficients
are gauge independent already. The result is that apart from the 
above example similar gauge cancellations occur in the 
case of all time ordered products which are related to
physical operators by CI. To summarize, as long as one
works with an overcomplete operator basis in which
some of the operators are not linearely independent, 
physical operators may have gauge dependent
coefficients. However, if one finally
projects onto to the physical operator basis,
these gauge artefacts must drop out.

We now come to the presentation of our results. 
The anomalous dimensions corresponding to the physical operator basis
can be cast into block diagonal form
\begin{equation}
\label{eq:block}
  \hat\gamma^{(3)} =
    \begin{matrix}
      \vphantom{{\displaystyle\int}}
        & \qquad\begin{matrix}
              \vec{\mathcal{O}}^{(3)}
            & \vec{\mathcal{T}}^{(12)}
            & \vec{\mathcal{T}}^{(111)}
          \end{matrix} \\
      \begin{matrix}
        \vec{\mathcal{O}}^{(3)}\\
        \vec{\mathcal{T}}^{(12)}\\
        \vec{\mathcal{T}}^{(111)}\\
      \end{matrix} &
      \begin{pmatrix}
      \hat{\gamma}^{(3)}_l & 0 & 0 \\
    \hat{\gamma}^{(12)}_l & \hat{\gamma}^{(12)}_{nl} & 0\\
    \hat{\gamma}^{(111)}_l & \hat{\gamma}^{(111)}_{nl} & 
                           \hat{\gamma}^{(111)}_{nl}
      \end{pmatrix}
    \end{matrix}
\end{equation}
The first column of this matrix is the result of our calculation 
and is presented in appendix 
\ref{sec:app}. Weinbergs theorem \cite{Weinberg} guarantees, 
that only local operators
are needed as renormalization counterterms.
However, apart from mixing with local operators the triple insertions 
$\mathcal T^{(111)}_{ijk}$
also
require nonlocal renormalization by the double insertions
$\mathcal T^{(12)}_{ij}$.
This reflects the fact that two operator components of the 
triple insertions are themselves renormalized by a local counterterm
of dimension 6 (see equation (\ref{eq:nlren})). 
\section{Aspects of the calculation}
\label{sec:calculation}

\begin{figure}
\begin{center}
\leavevmode
\epsfxsize=9cm
\epsffile[150 640 450 730]{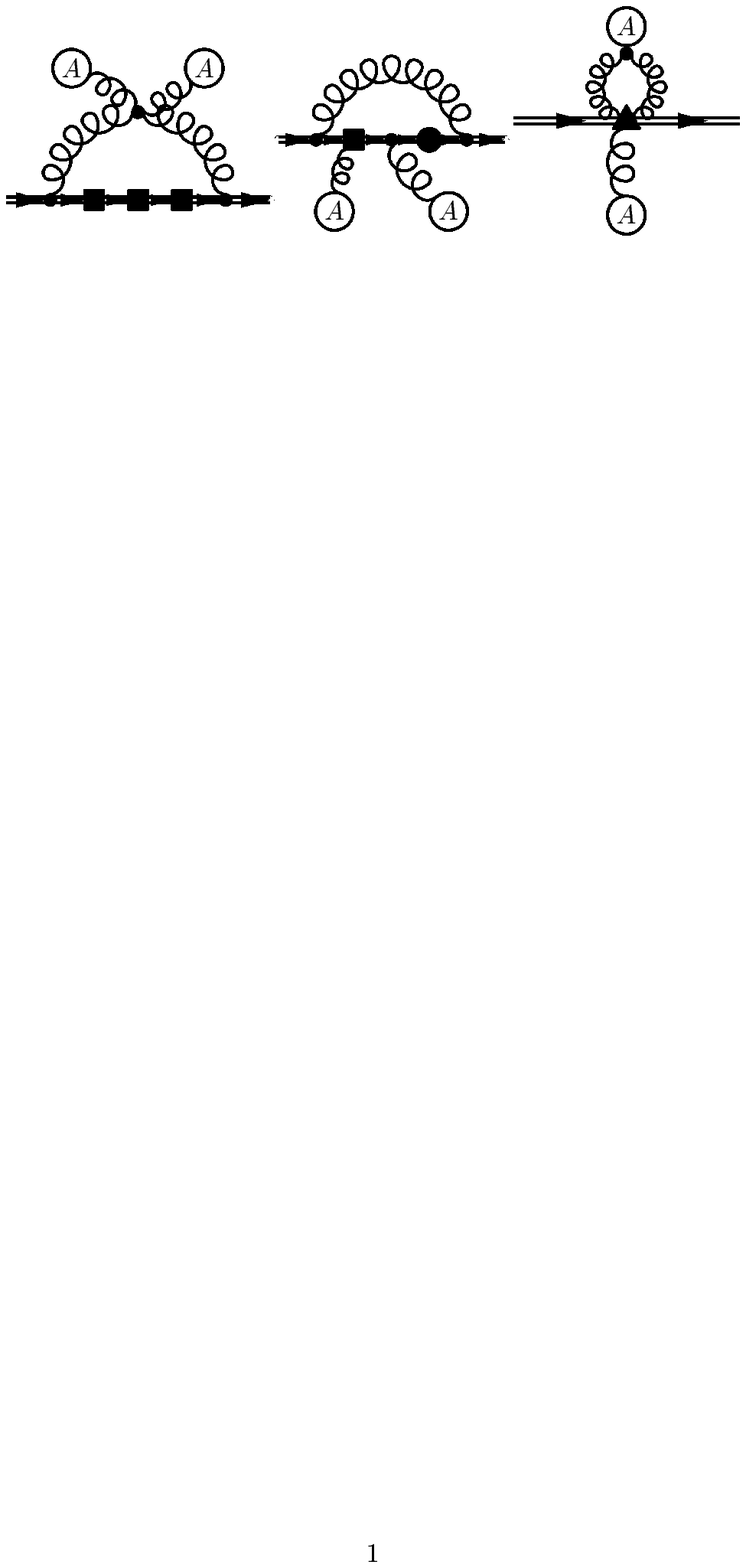} 
\caption{\label{fig:diag1} Examples of abelian and nonabelian diagrammatic
contributions to the renormalization of the 
triple, double and local insertion. 
The square, blob and triangle represent operators of dimensions
5, 6 and 7, respectively.}
\end{center}
\end{figure}

To calculate the anomalous dimensions of an operator basis
of definite dimensionality one generally has to insert every operator into 
potentially UV divergent 1PI Greenfunctions and extract 
the pole piece. In the background field method \cite{BFM,BFM/operators}, gauge 
invariance is maintained explicitly, which allows 
one to get the full result from the calculation of a
subset of Feyman diagrams with a specified number of
external background fields. 
This number is fixed by 
the highest power of gluon field strength tensors
appearing in the basis, because such operators 
do  not contribute at tree level to 1PI Greensfunctions
with a smaller number of external background fields.  
In our case the local operators $\mathcal O^{(3)}_{12/13}$ are 
bilinear in the field strength. This in turn
requires the calculation of all one loop diagrams 
with one incoming and outgoing heavy quark and two external background 
fields to get the 
counterterm contributions of all operators.
Figure \ref{fig:diag1} shows some diagrammatic examples.
The calculation has been performed with the
help of algebraic manipulations provided by 
the package FORM \cite{form}. 

In the first step every integrand corresponding to 
a specific diagram is expanded in a Taylor series
with respect to its external momenta, i.e. the
heavy quark momentum and two background field momenta. This
procedure is justified since we are only interested
in  UV divergences. The latter are easily extracted 
at this order of the expansion where the power
of external momenta agrees with the UV degree
of divergence of the diagrams. They appear in the
pole parts of simple one loop tensor integrals of  
generic structure,
\begin{equation}
I_{n,m}=\int \frac{d^{d}k}{(2\pi)^{d}}\frac{k^{\mu_{1}}
     \cdots k^{\mu_{n}}}{(vk)^{m}(k^{2})^{(n+4-m)/2}}\,, 
\end{equation}
which can be related recursively to two basis integrals
$I_{0,0}$ and $I_{0,2}$ via Pasarino--Veltman reduction. 
In the calculation integrals
with up to 10 indices are needed.

All diagrammatic contributions to a specific
operator insertion are then summed up and expressed 
in terms of tree level contributions
of the local operators $\mathcal O^{(3)}_{i}$:
\begin{equation}
\label{eq:opexp}
\langle \mathcal A_{i} \rangle^{(1)}_{1PI}=
(\frac{\alpha}{\pi})\frac{1}{\epsilon}\sum_{i=0}^{13}\gamma_{ij}
     \langle \mathcal O^{(3)}_{i} \rangle^{(0)}_{1PI}
\end{equation}
Here $\mathcal A_i$ denotes collectively local operators
or time ordered products of the full operator basis and $\langle \ldots 
\rangle_{1PI}$
indicates insertion of the given operator in a 1PI Greensfunction
to the loop order specified by the superscript. 

In this step explicit gauge invariance in the background
field ensures that no gauge variant operators
are needed in the series on the right hand side
of (\ref{eq:opexp}). This property provides a powerful
consistency check of the calculation, since
the tree level contributions contain up to 30 terms
which must match  in such a way that all
gauge variant terms cancel. 

In (\ref{eq:opexp}) the coefficients $\gamma_{ij}$ are 
easily identified as the entries of the anomalous dimension
matrix $\hat{\gamma}^{(3)}_l$,
    $\hat{\gamma}^{(12)}_l$ and $\hat{\gamma}^{(111)}_l$ corresponding 
to mixing with the local operators $\mathcal O^{(3)}_{i}$.
In the case of time ordered products the anomalous dimensions
of their operator components  contribute just additively and can 
therefore be derived
from the anomalous dimensions of the operator bases appearing at lower
$\mathcal O(1/m_Q)$. 

However, up to now the anomalous dimensions correspond to the 
full operator basis and some artificial gauge dependencies of
physical operators show up. Therefore in the last step of the 
calculation the projection onto the physical operator basis
has to be performed. This is where 
unphysical gauge artefacts drop  out and one is left with
the gauge independent anomalous dimension matrix of  
the physical operator basis. These nontrivial
cancellations provide another powerful cross--check
of our results.

\section{Renormalization group logarithms}
\label{sec:RG-flow}

With the one loop anomalous dimensions and the tree level
matching coefficients we are now in the position
to solve the renormalization group (RG) equation for
the Wilson coefficients of the 
effective lagrangian at $\mathcal O(1/m_Q^3)$:
\begin{equation}
\label{eq:RG}
\frac{d}{d\ln \mu}\vec C^{(3)}(\mu) + \hat{\gamma}^{(3)\top} \vec C^{(3)}(\mu) = 0
\end{equation}
Here $\vec C^{(3)}(\mu)$ denotes the coefficents of the
physical operators $(\vec{\mathcal O}^{(3)},\vec{\mathcal T}^{(12)},
\vec{\mathcal T}^{(111)}$).
However, an analytical diagonalization of $\hat{\gamma}^{(3)}$
seems to be difficult. Instead we restrict
ourselves to the calculation of  the first logarithmic
correction $\propto \alpha_s \ln(\mu/m_Q)$ in 
the coefficients $\vec C^{(3)}(\mu)$.
The exact solution of (\ref{eq:RG}) reads
\begin{equation}
\label{eq:exact}
\vec C^{(3)}(\mu) = \biggl( \frac{\alpha_s(\mu)}{\alpha_s(m_Q)}
                    \biggr)^{\frac{\hat{\gamma}^{(3)\top}}{2\beta^{(0)}}}
                    \vec C^{(3)}(m_Q)
\end{equation} 
with the one loop running coupling
\begin{equation}
\frac{\alpha_s(\mu)}{\alpha_s(m_Q)} = 1 - 2\beta^{(0)}
(\frac{\alpha_s(\mu)}{\pi})
\ln(\frac{\mu}{m_Q})
\end{equation}
where $\beta^{(0)} = (33-2n_f)/12$ in the presence of 
$n_f$ light flavours.
Expanding (\ref{eq:exact}) to first order in the strong coupling we get
\begin{equation}
\vec C^{(3)}(\mu) = \vec C^{(3)}(m_Q) 
           -(\frac{\alpha_{s}(\mu)}{\pi})\ln(\frac{\mu}{m_Q})
             \hat{\gamma}^{(3)\top}\vec C^{(3)}(m_Q)
           + \mathcal O((\frac{\alpha_{s}(\mu)}{\pi})^2).
\end{equation}
With our result $\hat{\gamma}^{(3)\top}$ and the tree level 
matching coefficients 
$\vec C^{(3)}(m_Q)$ the Wilson coefficients are 
easily calculated. The result is shown in the table below.
\begin{center}
\begin{tabular}{|c|c|c|} \hline
$C^{(3)}_i(\mu)$& tree level & 
coefficient of $(\alpha_{s}(\mu)/\pi)\ln(\mu/m_Q)$\\ \hline
1 & $2$ & $25/3\,C_A -23/3\,C_F$ \\ \hline
2 & $-1$ & $-1/2\,C_A $ \\ \hline 
3 & $-1$  & $4\,C_A +8/3\,C_F$ \\ \hline
4  & $1$ & $-17/6\,C_A $ \\ \hline
5  & $-2$ & $5/3\,C_A+ 8/3\,C_F$ \\ \hline
6 & $1$ & $-\,C_A$ \\ \hline
7 & $1$  & $-13/6\,C_A -8/3\,C_F$ \\ \hline
8 & $1$ & $-\,C_A $ \\ \hline
9 & $1$ & $-4\,C_A $ \\ \hline
10 & $-1$  & $9/2\,C_A $\\ \hline
11 & $-1$  & $9/2\,C_A $ \\ \hline
12 & $0$  & $1/12\,C_A$  \\ \hline
13& $0$ & $-1/3\,C_A$ \\ \hline
\end{tabular}
\end{center}
We only present 
the coefficients 
of local operators, since the coefficients of the
time ordered products are the products
of the coefficients of their operator components.
We note that there is a subtlety in the determination of
the tree level coefficients $\vec C^{(3)}(m_Q)$.
Again the CI forbid their naive extraction from  the
tree level lagrangian at $\mathcal O(1/m_Q^3)$
\begin{equation}
\mathcal L^{(3)} = \bar h_v iD_{\mu}(ivD)^2 iD^{\mu} h_v
                  - \bar h_v (ivD)^4 h_v
                  -\bar h_v i\sigma^{\mu\nu}iD_{\mu} (ivD)^2 iD_{\nu} h_v. 
\end{equation}
The reason is that some of the time ordered products
with nonvanishing tree level coefficients are related
to physical operators by CI. Since we are studying
the RG flow of the physical basis, we first have to 
reduce the tree level matrix elements of the full operator basis with the CI 
to the ones of the physical basis and then apply the RG 
controlled by the anomalous dimensions of the physical basis.
  
For example, the triple insertion $\mathcal T^{(111)}_{113}$
contributes at tree level with coefficient $-1$. However,
after application of CI (\ref{eq:CI111},\ref{eq:CI12}) this 
coefficient is attributed to the local operator 
$\mathcal O^{(3)}_2 = \bar h_v (iD)^2(iD)^2 h_v$
which does not appear in the tree level lagrangian.
The same occurs for all time ordered products
which contribute at tree  level and in addition are
affected by a CI. This explains
the nonvanishing tree level contributions
of coefficients of operators not appearing in the tree level lagrangian.
Note that the operators $\mathcal O^{(3)}_{12/13}$ are protected
from CI contributions because of their color structure.

As an phenomenologic application we consider 
the expansion of the physical mass $M_H$ of a heavy meson in the HQET
\begin{equation}
\label{eq:mass}
M_H = m_Q + \bar \Lambda - \langle\mathcal{L}_{int}(0)e^{i\int d^4x
\mathcal{L}_{int}(x)} \rangle 
\end{equation} 
with $\mathcal{L}_{int}$ given by the powercorrections on the 
right hand side of (\ref{eq:HQETlag}) and $\langle \cdots \rangle$
denoting the expectation value between heavy meson states in the 
heavy mass limit.
Expanding up to $\mathcal O(1/m_Q^3)$ we get
\begin{equation}
M_H = m_Q + \bar \Lambda - \frac{1}{2m_Q}M_H^{(1)} 
                     -\frac{1}{(2m_Q)^2} M_H^{(2)} 
                  -\frac{1}{(2m_Q)^3} M_H^{(3)}
\end{equation}
where
\begin{align}
\label{eq:M1}
M_H^{(1)}  &= \sum_{i = 1}^2 C^{(1)}_i(\mu)\langle \mathcal O^{(1)}_i\rangle \\
\label{eq:M2}
M_H^{(2)}  &= \sum_{i = 1}^2 C^{(2)}_i(\mu)\langle \mathcal O^{(2)}_i\rangle
             +\sum_{i\leq j=1}^2 C^{(1)}_i(\mu)C^{(1)}_j(\mu)
             \langle \mathcal T^{(11)}_{ij}\rangle\\
\label{eq:M3}
M_H^{(3)}  &= \sum_{i = 1}^{13} C^{(3)}_i(\mu)\langle \mathcal O^{(3)}_i\rangle
             +\sum_{i,j=1}^2 C^{(1)}_i(\mu)C^{(2)}_j(\mu)
             \langle \mathcal T^{(12)}_{ij}\rangle\\
           & + \sum_{i\leq j\leq k=1}^2 
              C^{(1)}_i(\mu)C^{(1)}_j(\mu)C^{(1)}_k(\mu)
             \langle \mathcal T^{(111)}_{ijk}\rangle\nonumber \,.
\end{align}
Since the coefficients $C^{(1/2)}_i(\mu)$ are allready known from
lower order calculations, with our result for the coefficients
$C^{(3)}_i(\mu)$ all short distance contributions up to 
$\mathcal O(1/(2m_Q)^3)$
are known. However the problem is the nonperturbative input,
i.e. the hadronic matrix elements on the right hand side of 
(\ref{eq:M1},\ref{eq:M2},\ref{eq:M3}) 
which cannot be calculated from first
principles.    
The hadronic matrixelements at $\mathcal O(1/m_Q)$ are
parametrized by the well known parameters $\lambda_1$ and
$\lambda_2$. At $\mathcal O(1/m_Q^2)$ 5 parameters are needed
and at  $\mathcal O(1/m_Q^3)$ a further proliferation 
of parameters takes place: here 21 parameters are needed.
\section{Conclusions}
\label{sec:conclusions}

We have presented a new calculation of 
the anomalous dimensions of the operator basis
appearing at $\mathcal O(1/m_Q^3)$ of the
HQET lagrangian. Local operators  as well
as all time ordered products -- composed of lower dimension
operators -- of dimension 7 contribute to the operator basis at this order.
We have shown that there exist non trivial
relations among some time ordered products
and local operators, so called contraction identities, 
which have to be properly taken
into account during renormalization. Otherwise
gauge independence of the physical operators is violated.

The calculation of the one loop anomalous dimensions
of the physical operators was the goal of this paper.
Supplemented by the tree level matching coefficients,
it was possible to extract the first logarithmic corrections
$\propto \alpha_s(\mu)\ln(\mu/m_Q)$ to the Wilson coefficents.
Of course, to get the non--logarithmic contributions
at $\mathcal O(\alpha_s)$, 
a one loop matching calculation has to be performed, 
which is beyond the scope of this work.

In \cite{manohar} the short distance 
coefficients of the $\mathcal O(1/m_Q^3)$ effective lagrangian
have been derived by matching the full QCD vertex function with
one external gluon to HQET in an on--shell renormalization
scheme. In this case all loop diagrams in the
effective theory vanish and the coefficients
can be read of directly from an expansion
of the QCD vertex in powers of the heavy quark momentum and the
gluon momenta to the appropriate order. It is clear that the coeffients 
of operators bilinear in the gluon fields 
cannot be determined by this method.
Using the equation of motion we have transformed the operator basis
of \cite{manohar} into ours. This way only the coefficients of
$\mathcal O^{(3)}_2$, $\mathcal O^{(3)}_6$ and $\mathcal O^{(3)}_8$
are determined. The other coefficients are related to coefficients 
which have not been calculated in \cite{manohar}.
Unfortunately there is a mismatch in the coefficient of $\mathcal O^{(3)}_2$.
A possible reason for this mismatch may be 
found in the fact that the tensor structures appearing in the
expansion of the one loop QCD vertex function at $\mathcal O(1/m_{Q}^{3})$
cannot be uniquely attributed to tree level insertions of the local operators 
$\mathcal O^{(3)}_{i}$ with one external background field. 
For example, in our conventions 
the operators $\mathcal O^{(3)}_{3}$ and $\mathcal O^{(3)}_{4}$ 
contribute at tree level identically  to the vertex function
with only one external gluon.
Therefore 
one should expand the one loop QCD vertex function with two
external background fields in inverse powers of the large mass  
to  perform the matching at $\mathcal O(1/m_Q^3)$ correctly.

In \cite{bauer} the effect of four fermion operators
at $\mathcal O(1/m_Q^2)$ has been studied. 
The upshot of this work was that the famous Darwin
operator is renormalized by a four fermion
operator involving two static and two light quarks,
even if all four fermion operators are removed at the end of the 
renormalization procedure.
Similiar effects cannot be excluded at $\mathcal O(1/m_Q^3)$,
but in this case the higher dimension of the operators 
make life more complicated. Nevertheless it 
is not only a calculational challenge
to include all possible four fermion operators
into the operatorbasis at $\mathcal O(1/m_Q^3)$
and to investigate their renormalization properties \cite{future}.

Another topic which will be covered in a sequel to this
publication \cite{Future}, is the reparametrization invariance of HQET.
This new symmetry of the HQET lagrangian predicts
relations among Wilson coefficients of operators 
appearing at different  $\mathcal O(1/m_Q)$.
With our results it should be  possible to check the 
validity of these relations. This in turn may help
to draw a decision between several
concepts of this symmetry appearing in the
literature, all of them unfortunately agreeing
in their predictions up to $\mathcal O(1/m_Q^2)$. 

\section*{Acknowledgements}
This work is supported by DFG contract Ma 1187/7-1,2 and 
Graduiertenkolleg ``Elementarteilchenphysik an Beschleunigern''.
The author wish to thank T. Mannel for useful and enlightening discussions.

\appendix
\section{Anomalous dimension matrices}
\label{sec:app}
In the decomposition $\hat{\gamma}^{(\cdots)}_l =C_A \hat{\gamma}^{(\cdots)A}_l
+C_F\hat{\gamma}^{(\cdots)F}_l$ the entries in the first column of
(\ref{eq:block}) are:
\begin{eqnarray*}
\hat{\gamma}_l^{(3)A} =\quad\quad\quad\quad\quad\quad\quad\quad\quad\quad\quad\quad\quad\quad\quad\quad\quad\quad\quad\quad\quad\quad\quad\quad\quad\quad\quad\quad\quad\quad\quad\quad\nonumber\\ \left(
\begin{array}{ccccccccccccc}
-\frac{11}{12} &0 &\frac{11}{24} &-\frac{11}{24} &0 &0 &0 &0 &0 &0 &0 &\frac{11}{288} &-\frac{11}{72} \\
-4 &0 &1 &-1 &0 &0 &0 &0 &0 &0 &0 &0 &-\frac{2}{9} \\
-3 &\frac{5}{6} &\frac{5}{6} &-\frac{5}{3} &0 &0 &0 &0 &0 &0 &0 &\frac{1}{12} &-\frac{7}{18} \\
-\frac{7}{2} &\frac{5}{6} &\frac{4}{3} &-\frac{13}{6} &0 &0 &0 &0 &0 &0 &0 &\frac{1}{24} &-\frac{11}{36} \\
0 &0 &0 &0 &-\frac{7}{6} &\frac{1}{8} &\frac{1}{12} &\frac{1}{8} &\frac{1}{3} &-\frac{1}{3} &-\frac{1}{3} &0 &0 \\
0 &0 &0 &0 &0 &-\frac{7}{8} &\frac{1}{2} &\frac{5}{24} &\frac{1}{3} &-\frac{1}{4} &-\frac{5}{12} &0 &0 \\
0 &0 &0 &0 &\frac{1}{2} &\frac{5}{24} &-1 &\frac{5}{24} &-\frac{1}{12} &\frac{1}{12} &\frac{1}{12} &0 &0 \\
0 &0 &0 &0 &0 &\frac{5}{24} &\frac{1}{2} &-\frac{7}{8} &\frac{1}{3} &-\frac{5}{12} &-\frac{1}{4} &0 &0 \\
0 &0 &0 &0 &1 &\frac{1}{4} &0 &\frac{1}{8} &0 &-\frac{1}{2} &-\frac{1}{2} &0 &0 \\
0 &0 &0 &0 &\frac{1}{2} &\frac{1}{24} &\frac{1}{4} &\frac{1}{6} &\frac{11}{24} &-\frac{13}{12} &-\frac{1}{3} &0 &0 \\
0 &0 &0 &0 &\frac{1}{2} &\frac{1}{6} &\frac{1}{4} &\frac{1}{24} &\frac{11}{24} &-\frac{1}{3} &-\frac{13}{12} &0 &0 \\ 
0 &0 &0 &0 &0 &0 &0 &0 &0 &0 &0 &0 &0 \\
0 &0 &0 &0 &0 &0 &0 &0 &0 &0 &0 &\frac{11}{24} &- \frac{1}{6} 
\end{array}\right)\\
\hat{\gamma}_l^{(111)A} = \left(
\begin{array}{ccccccccccccc}
\frac{16}{3}&\frac{1}{6}&-\frac{5}{2}&\frac{7}{3}&0&0&0&0&0&0&0&0&0
\\
0&0&0&0&-\frac{4}{3}&\frac{5}{3}&\frac{8}{3}&\frac{5}{3}&3&-4&-4&0&0
\\
5&\frac{1}{6}&-\frac{23}{6}&\frac{11}{3}&-1&0&1&0&\frac{5}{2}&-\frac{5}{2}&-\frac{5}{2}&-\frac{1}{4}&\frac{1}{2}
\\
0&0&0&0&0&0&0&0&0&0&0&\frac{1}{8}&\frac{1}{4}
\\
\end{array}\right)\\
\hat{\gamma}_l^{(111)F} = \left(
\begin{array}{ccccccccccccc}
\frac{32}{3}&\frac{8}{3}&-\frac{8}{3}&0&0&0&0&0&0&0&0&0&0
\\
0&0&0&0&\frac{8}{3}&-\frac{4}{3}&\frac{8}{3}&-\frac{4}{3}&0&0&0&0&0
\\
0&0&0&0&0&0&0&0&0&0&0&0&0 \\
0&0&0&0&0&0&0&0&0&0&0&0&0
\\
\end{array}\right)\\[2cm]
\hat{\gamma}_l^{(12)A} =\quad\quad\quad\quad\quad\quad\quad\quad\quad\quad\quad\quad\quad\quad\quad\quad\quad\quad\quad\quad\quad\quad\quad\quad\quad\quad\quad\quad\quad\quad\quad\quad\nonumber\\ \left(
\begin{array}{ccccccccccccc}
4&-\frac{1}{2}&-\frac{3}{2}&2&0&0&0&0&0&0&0&0&\frac{2}{9}\\
0&0&0&0&-\frac{4}{3}&-\frac{7}{12}&-1&-\frac{7}{12}&-\frac{1}{6}&\frac{2}{3}&\frac{2}{3}&0&0\\
0&0&0&0&0&\frac{1}{24}&\frac{1}{2}&\frac{1}{24}&\frac{7}{12}&-\frac{7}{12}&-\frac{7}{12}&0&0\\
-\frac{2}{3}&\frac{1}{3}&\frac{17}{12}&-\frac{7}{4}&\frac{1}{6}&-\frac{5}{24}
&-\frac{1}{3}&-\frac{5}{24}&-\frac{3}{4}&\frac{3}{4}&\frac{3}{4}&\frac{1}{144}&\frac{11}{36}\\
\end{array}\right)\\
\hat{\gamma}_l^{(12)F} = \left(
\begin{array}{ccccccccccccc}
0&0&0&0&0&0&0&0&0&0&0&0&0\\
0&0&0&0&-\frac{16}{3}&0&0&0&0&0&0&0&0\\
0&0&0&0&0&0&0&0&0&0&0&0&0\\
0&0&0&0&0&0&0&0&0&0&0&0&0\\
\end{array}\right)
\end{eqnarray*}
$\hat{\gamma}_l^{(3)F}$ has only zero entries, a fact for which
we have no simple explanation by now.
 

\end{document}